ASTRONOMY

# Polarized accretion shocks from the cosmic web

Tessa Vernstrom[1,2]*, Jennifer West[3], Franco Vazza[4,5,6], Denis Wittor[6,4], Christopher John Riseley[4,5,2], George Heald[2]

On the largest scales, galaxies are pulled together by gravity to form clusters, which are connected by filaments making a web-like pattern. Radio emission is predicted from this cosmic web, which should originate from the strong accretion shocks around the cosmic structures. We present the first observational evidence that Fermi-type acceleration from strong shocks surrounding the filaments of the cosmic web, as well as in peripherals of low-mass clusters, is at work in the Universe. Using all-sky radio maps and stacking on clusters and filaments, we have detected the polarization signature of the synchrotron emission with polarization fractions ≥20%, which is best explained by the organization of local magnetic fields by strong shock waves both at the cluster peripheries and between clusters. Our interpretation is well supported by a detailed comparison with state-of-the-art cosmological simulations.

## INTRODUCTION

The cosmic web is the term used to refer to the clusters (and superclusters), filaments, and voids that make up the large-scale structure of the Universe (1). In Λ cold dark matter (ΛCDM) cosmology, this web is formed from the anisotropic gravitational collapse of matter from primordial overdensities. Clusters of galaxies are the most dense regions of space, forming at the intersections of the cosmic web. Galaxy clusters are connected by filaments. However, voids, the most underdense regions in the Universe, fill most of the volume. Since the seminal works by Y. B. Zeldovich in 1970 (2), it has been understood that the formation of self-gravitating cosmic structures cannot proceed without having ordinary matter undergo strong shock heating, which must fill enormous volumes of the Universe with $T \sim 10^5$ to $10^7$ K of plasma.

The synchrotron cosmic web refers to the synchrotron emission that is expected to be generated in these regions from intergalactic magnetic fields and by electrons accelerated by shocks from infalling matter through the diffusive shock (Fermi-I) acceleration mechanism (DSA) (3). Many simulations predict and model this synchrotron emission (4–9); however, observational confirmation of these models has remained elusive.

There have been numerous detections and studies of diffuse radio emission in clusters (10–12). This emission consists of different categories such as radio halos, which are centrally located large-scale diffuse sources that roughly follow the intracluster medium mass distribution and are generally produced through second-order Fermi acceleration (Fermi-II) (13, 14) and/or secondary electrons for cosmic ray acceleration (15–17). Another class of diffuse emission is that of cluster shocks or relics. These are diffuse sources that come from particles (re)accelerated by intracluster shock waves from mergers (18–20). Of these types of sources, only about 100 have been detected, and they have only been seen in high-mass clusters (mass in solar mass units $\geq 10^{14}$ $M_\odot$). More than 40% of clusters show evidence of a disturbed dynamical state (mergers) (21), while less than 10% have had detections of radio emission that are expected to accompany mergers (10).

It is thought that similar processes should be present in between clusters in intercluster bridges and filaments as well. To date, only a few inter- and intracluster bridges have been found to host detectable diffuse radio emission (22–25). Even within the now most well-studied radio intercluster bridge, the particle acceleration mechanism for the emission is debated between Fermi-I processes (22) and more turbulent Fermi-II processes (26). While such radio detection is compatible with synchrotron emission by weak magnetic fields organized on scales of megaparsecs, the mechanism that could accelerate ultrarelativistic electrons in such diluted plasmas (or the potential contribution from radio galaxies or other sources) remains more debatable. A plausible scenario to refill such giant volumes, away from any other galaxy, would be Fermi-type acceleration by shocks forming at the periphery of filaments. The existence of such shocks is still an unproven core concept of our models of cosmic structure formation and growth. The magnetic field strength in filaments is still not well known, with values from diffuse synchrotron emission ranging from 30 to 100 nG (from either equipartition or inverse Compton scattering calculations or comparison with DSA simulations) (27–30) and from Faraday rotation measure (RM) studies ranging from 10 to 100 nG (31–35). Moreover, it is still unknown how turbulent the environment inside filaments is.

While the overdense matter distribution tracked by galaxies has long since been detected with optical and infrared surveys, the presence of diluted and hot plasmas in between galaxies has so far been detected only in a statistical way. The method of stacking filaments, or pairs of clusters to detect filamentary emission between them, has had success in recent years via the thermal Sunyaev-Zel'dovich effect (36, 37), weak lensing (38, 39), the dark matter mass-to-light ratio (40), thermal x-ray emission (27, 41, 42), and, by our group, radio synchrotron emission (27). However, in general, the number of known clusters is not large enough for such stacking experiments, and a tracer for clusters and groups must be used. Pairs of luminous red galaxies (LRGs) are a commonly used tracer. LRGs are early-type massive galaxies that are known to usually reside in the centers of groups or clusters (*43–45*) and are identified in optical surveys in much larger numbers than galaxy clusters. Stacking many clusters or pairs of clusters decreases the noise, allowing one to look for an average signal from the objects well below the noise level of the image. In some cases, stacking may be the only method of observing such faint or diffuse emission (as direct imaging, even with future telescopes, may still have imaging limitations and be affected by confusion of galaxies).

[1]ICRAR, The University of Western Australia, 35 Stirling Hw, 6009 Crawley, Australia. [2]CSIRO Space & Astronomy, PO Box 1130, Bentley, WA 6102, Australia. [3]Dunlap Institute for Astronomy and Astrophysics, University of Toronto, Toronto, ON M5S 3H4, Canada. [4]Dipartimento di Fisica e Astronomia, Università degli Studi di Bologna, via P. Gobetti 93/2, 40129 Bologna, Italy. [5]INAF–Istituto di Radioastronomia, via P. Gobetti 101, 40129 Bologna, Italy. [6]Hamburger Sternwarte, Gojenbergsweg 112, 21029 Hamburg, Germany.
*Corresponding author. Email: tessa.vernstrom@uwa.edu.au

Using LRGs from the Sloan Digital Sky Survey (SDSS) Data Release 7 LRG catalog (*46*), we stacked 612,025 pairs of LRGs or clusters that are physically near each other in three-dimensional (3D) space, i.e., with physical separations between 1 and 15 Mpc. These cluster pairs are most likely connected by filaments (from here, these are referred to as the "connected" pairs, although not all of the pairs are in reality connected by filaments). To perform the stacking, cutouts were made of all the pairs, which were then rotated and scaled. This ensured that the clusters and filaments would line up and any signal would add together. To compare, we also stacked on pairs of clusters that are separated in physical space by hundreds or thousands of megaparsecs although, due to projection, still appear near to each other on the sky (these pairs are a control sample where strong filamentary emission is not expected and are referred to as "unconnected" pairs). We stacked on data from two radio surveys: the 1.4-GHz total and polarized intensity maps from the recently released High-band north (HBN) Global Magneto-Ionic Medium Survey (GMIMS) survey (*47*) and the 30-GHz maps in total and polarized intensity from the Planck mission (*48*). In addition to stacking on pairs of clusters, we also stacked on single clusters, also using the LRGs as cluster tracers, to look for diffuse polarized emission from these low-mass ($\leq 10^{14} M_\odot$) groups and clusters. Using the polarized intensity data is key for determining the type of emission processes at play in any detected signal, as well as for providing information about the ordering of the magnetic fields or turbulence of the environment.

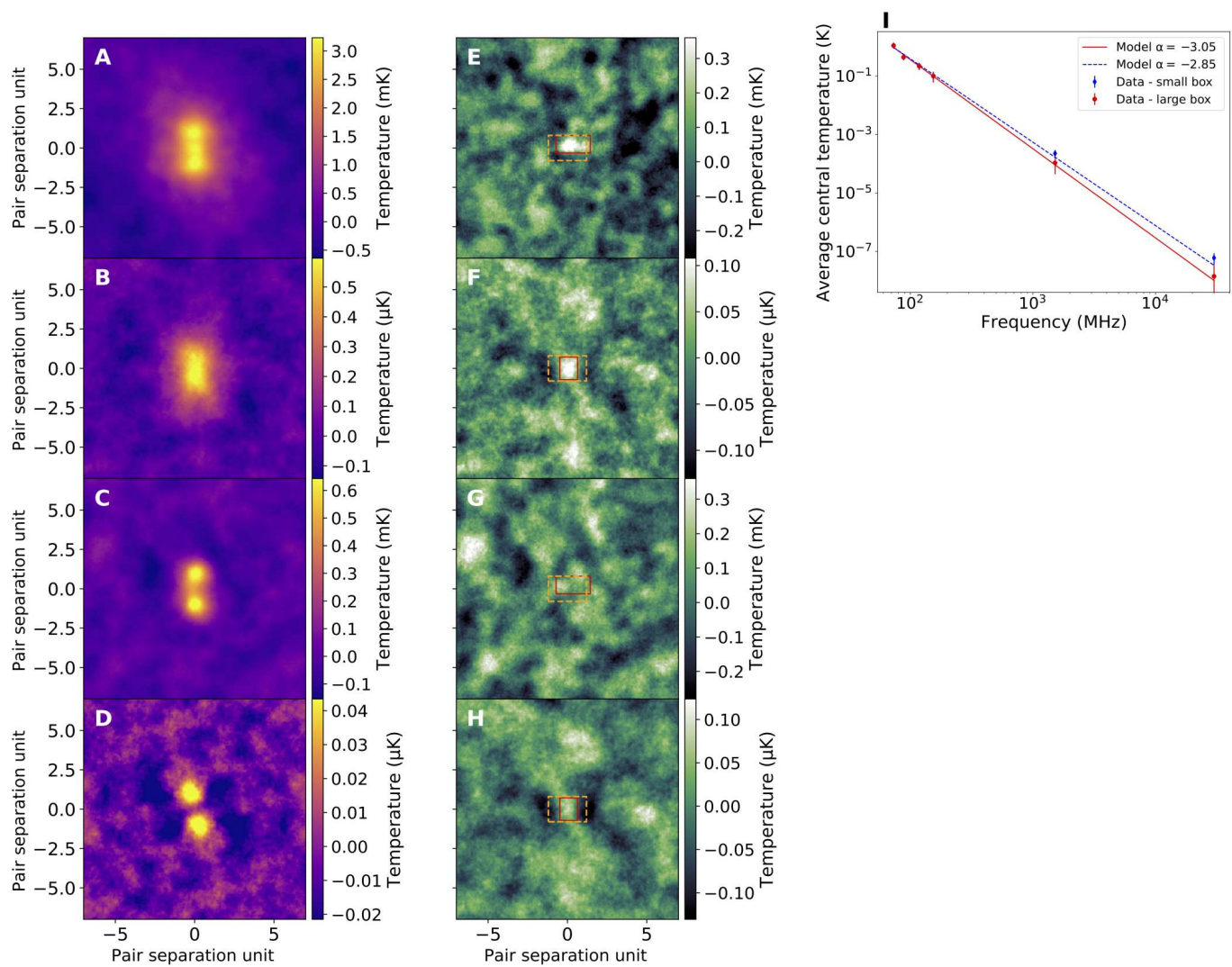

**Fig. 1. Stacked images and residual images from the Stokes *I* total intensity maps.** The first column (**A** to **D**) is the stacked images, whereas the second column (**E** to **H**) is the residual after model subtraction. Top to bottom: The rows are connected pairs of 1.4 GHz, connected pairs of 30 GHz, unconnected pairs of 1.4 GHz, and unconnected pairs of 30 GHz. The boxes show the area used to measure the average signals, with the yellow being from the low-frequency work and the red boxes being sized to the total intensity signal in this work. (**I**) Spectrum from the four previously detected low-frequency data points (*27*) and the two high-frequency points for the detected intercluster emission, with the red points and line measured from inside the yellow box of (E) and (F) and the blue points measured from the smaller red boxes.

**RESULTS**

In all cases of cluster pair stacking, we produced an image of the cluster pairs. We then modeled the cluster emission (excluding the central intercluster region), and we subtracted it to obtain a residual image showing any excess emission between the clusters that could be attributed to filaments or intercluster bridges.

In the sample of connected cluster pairs, we detect intercluster emission in total intensity at both frequencies between the clusters. This emission is not detected in the sample of unconnected cluster pairs. The mean values for the intercluster regions are measured inside of a box whose height is set to be just less than the separation of the LRG positions, with the separations ranging from 1 to 15 Mpc, and the width set to include the largest detected signal in all the maps. The size does not correspond to an exact physical scale or size, but rather the pixel sizes are set relative to the cluster pair separation. Statistics were computed over all the pixels in this box for all the stack residual images of intercluster regions in this work and in the previous low-frequency detections (27) to compare the values over the same regions. The detected signal in both polarization maps and all low-frequency maps is well fit by the same box size; however, both the total intensity residuals at 1.4 and 30 GHz are spatially smaller than the polarized signals; thus, computing the mean in this box for the total intensity may underestimate the signal as it includes more pixels without detectable emission. It is not clear whether this is a physical effect, e.g., the total intensity is more concentrated in the inner region of a filament, or decreased signal-to-noise ratio of the total intensity across the region, or some combination. However, because of this, we also fit smaller boxes separately to the two total intensity residual signals using a curve of growth to choose the box size for each signal. We report the averages in these smaller boxes, as well as the bigger box used in the low-frequency work, which is more fitted to the polarized emission as well. The images from the total intensity stacking, with the different box sizes highlighted, and the spectrum from the intercluster emission are shown in Fig. 1.

For the brightness temperature spectral index when combined with the low-frequency radio detections (27), over the range of 73 MHz to 30 GHz, we find $\alpha_T = -3.05 \pm 0.15$ when using the values from the larger low-frequency box and $\alpha_T = -2.85 \pm 0.1$ when using the smaller total intensity boxes. Converting from brightness temperature to flux density, the spectral index $\alpha = -1.05 \pm 0.15$ and $\alpha = -0.85 \pm 0.1$. These are consistent with each other to within uncertainties. This spectral index is consistent with the expectation of electron acceleration by quasi-stationary strong shocks. Such strong shocks are similar to the structure formation shocks that are expected at the extreme periphery of cosmic structures. If we assume that the emission is from shocks, we can use the spectral index to get the average Mach number of the shocks, $\mathcal{M}$

$$\mathcal{M} = \sqrt{\frac{1-\alpha}{-1-\alpha}} \quad (1)$$

In this case, the Mach number for the signal in the intercluster region is $\mathcal{M} = 6.4$ assuming $\alpha = -1.05$, with Mach numbers as low as 3.3 with the 1σ uncertainty in the spectral index (the Mach number calculation breaks down for $\alpha > -1$).

We also find an even stronger detection of polarized emission, at both frequencies, in between connected clusters. No significant detection was found for the unconnected pairs. The polarized fraction (polarized intensity divided by total intensity) of this filamentary emission between the connected clusters is found to be ~12 to 50% at both frequencies at the peaks of the total intensity emission, with values of 20 to 60% when comparing the averages over the same regions (see Table 1). This is considered a high polarization fraction as normal galaxies are in the low percent range, and this implies the presence of ordered magnetic fields. The images from the pair stacking in polarized intensity are shown in Fig. 2. The values for the measured residual signals in the region between cluster pairs in all connected cases are reported in Table 1 (with the values for the unconnected or null sample given in table S2). We note that a two-sample Kolmogorov-Smirnov test was performed between the connected and unconnected pairs from a distribution of intercluster region values for both frequencies and polarizations, and all had $P \ll 0.00001$.

When looking at single clusters, if there are low-level shocks or relics, these would be situated at the periphery and again have high polarization fractions, with the center of the clusters expected to be depolarized due to turbulence. At both frequencies, we find bright emission in total intensity peaking in the center of the clusters and falling off radially to a distance of approximately 1 Mpc. In the polarized intensity, the emission extends further out to a distance of 2 to 3 Mpc. The brightness in the 1.4-GHz polarized intensity falls off radially as well but more slowly than in the total intensity. The 30-GHz polarized intensity is strongly depolarized in the central region out to a distance of roughly 0.5 Mpc, with a brighter ring of emission outside the center. Looking at the polarized fraction of the 1.4-GHz single clusters, as opposed to the polarized intensity, the depolarization in the center is much more apparent. The polarized fraction shows a ring of higher polarization fraction around the cluster center. The polarization fraction near the cluster center is ≤10% increasing to 20 to 25% in the ring. If we fit a Gaussian model to the 1.4-GHz polarized intensity, based on the width and shape outside the center, we can see that the center is about 10% depolarized compared to the model Gaussian. The images of the single cluster stacks can be seen in Fig. 3. The polarization fractions from the full pair images, the pair residual images, and single cluster are shown in Fig. 4.

**Table 1. Average values for the central residuals of the stacked images after model subtraction.** *F* is the polarized fraction. The "big box" values refer to the measurement of the mean inside the yellow box as seen in Figs. 1 and 2 and the same across all maps and those in (27). The "small box" values refer to the means measured in smaller boxes optimized for the total intensity signal, separately in the GMIMS and Planck data (shown as red boxes in Figs. 1 and 2. The uncertainties listed are detailed in the "Uncertainty" section in Materials and Methods. The uncertainties in parentheses for the polarization fractions are the uncertainties when taking into account the covariance. The $T_{peak}$ refers to the peak value for each map in the central regions, whereas Peak(*I*) gives the total intensity peak values and the polarized intensity at those total intensity peak locations. Similar values for the unconnected or control sample are shown in table S2.

| Name | $\langle T_{big\ box}\rangle_{connected}$ | $\langle T_{small\ box}\rangle_{connected}$ | Peak $T_{connected}$ | Peak(*I*) $T_{connected}$ |
|---|---|---|---|---|
| GMIMS *I* (mK) | 0.11 ± 0.06 | 0.22 ± 0.07 | 0.6 ± 0.1 | 0.6 ± 0.1 |
| GMIMS *P* (mK) | 0.09 ± 0.03 | 0.14 ± 0.05 | 0.31 ± 0.06 | 0.30 ± 0.06 |
| GMIMS *F* (%) | 80 ± 50 (40) | 60 ± 30 (23) | 50 ± 10 (10) | 50 ± 10 (10) |
| Planck *I* (μK) | 0.02 ± 0.02 | 0.08 ± 0.02 | 0.14 ± 0.04 | 0.14 ± 0.04 |
| Planck *P* (μK) | 0.015 ± 0.004 | 0.018 ± 0.005 | 0.038 ± 0.008 | 0.018 ± 0.008 |
| Planck *F* (%) | 75 ± 80 (70) | 22 ± 8 (6) | 20 ± 8 (7) | 12 ± 6 (5) |

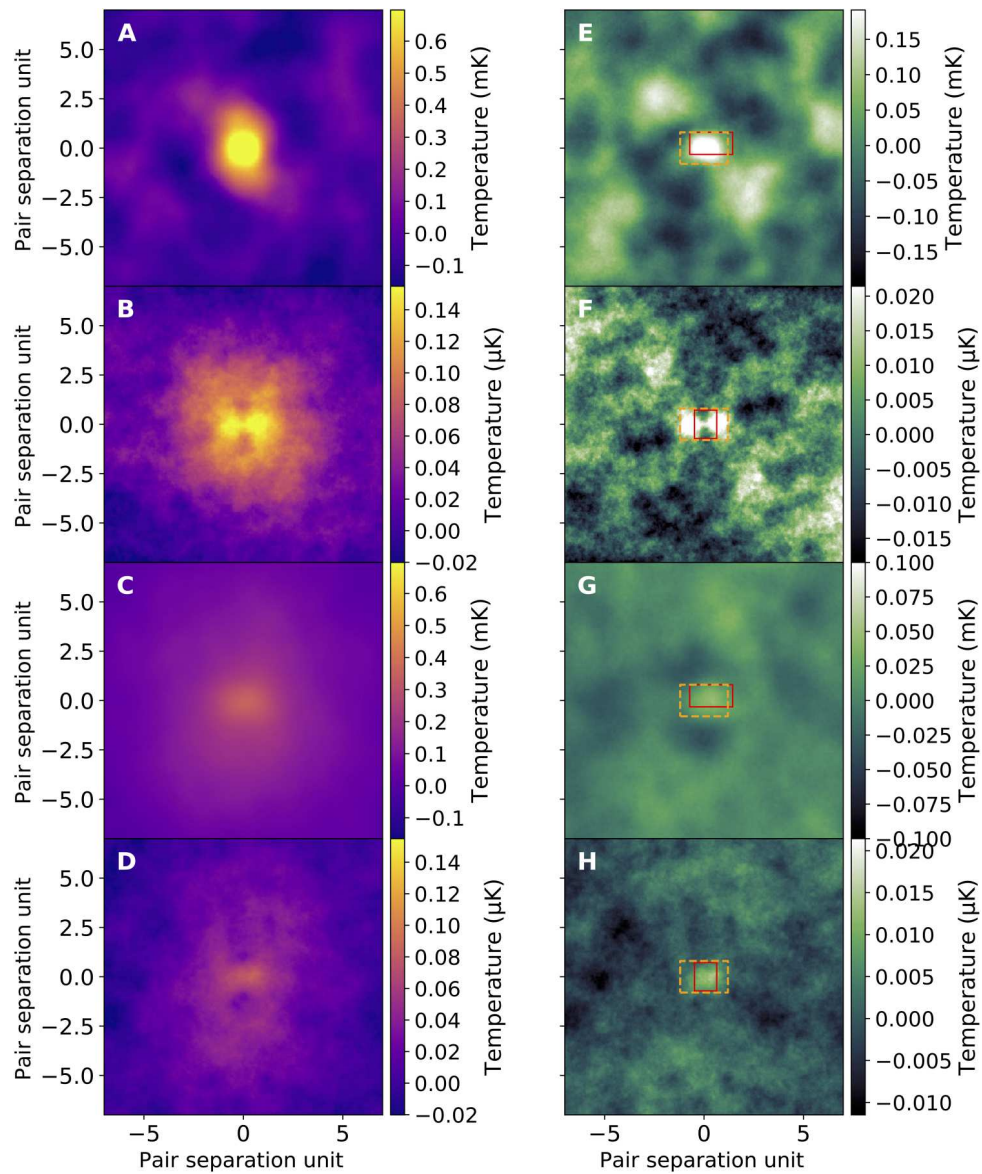

**Fig. 2. Stacked images and residual images from the polarized intensity maps.** The first column (**A** to **D**) is the stacked images, whereas the second column (**E** to **H**) is the residual after model subtraction. Top to bottom: The columns are connected pairs of 1.4 GHz, connected pairs of 30 GHz, unconnected pairs of 1.4 GHz, and unconnected pairs of 30 GHz.

The ring or offset nature of the polarized emission around the clusters explains the difference in appearance between the pair images in total and polarized intensity (see Figs. 1 and 2). In the case of the 1.4-GHz polarization data, a single central bright source is seen in the pairs image, rather than the two bright sources at the positions of the clusters, as is seen in the total intensity image. In the 30-GHz polarized intensity pairs image, there is a centrally located bright emission, and at the positions of the two clusters, there appears to be a deficit of signal. By comparing these images of the connected pairs with the simple addition of stacking of single clusters and combining them, we find no excess between the unconnected pairs, whereas the connected pairs still show a detection in between the clusters. Further details about the uncertainties in the reported values are given in Materials and Methods.

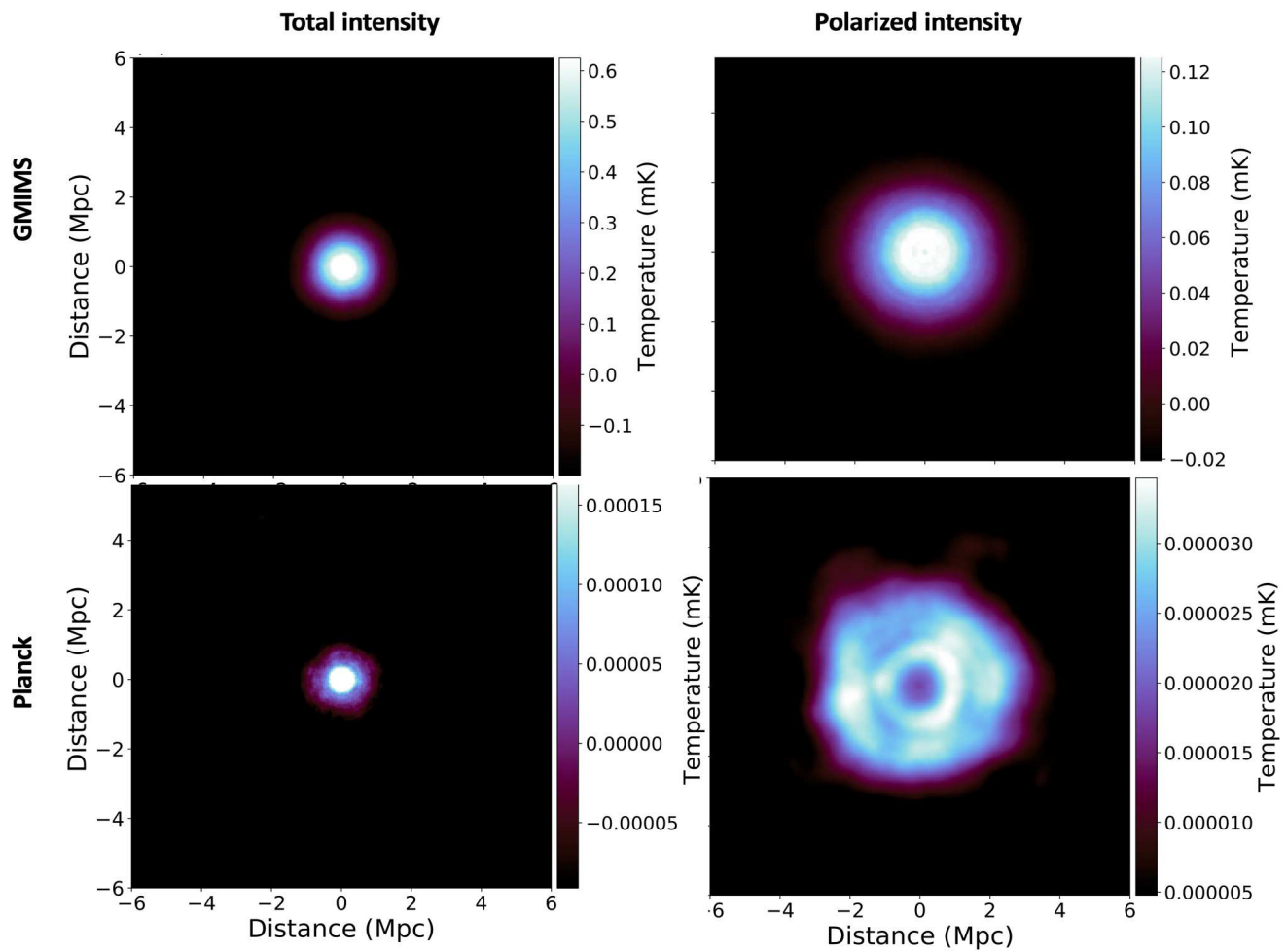

**Fig. 3. Stacks on single clusters.** The top row shows the 1.4-GHz images; total intensity is on the left, and polarized intensity is on the right. The bottom row shows the 30-GHz Planck maps; again, total and polarized intensities are on the left and right, respectively.

### DISCUSSION
To make sure that the detection is really related to clusters and not just chance or random emission detected on the sky, we repeated the detection stacking experiment using random sky positions instead of clusters, and not once was any signal seen in neither total nor polarized intensity. We also stacked on pairs of known radio galaxies instead of clusters and found the spectral index in total intensity to be much shallower ($α_T = -2.4$). No detection was seen in polarized intensity at 1.4 GHz, and at 30 GHz, only a slight polarized signal was seen at the location of the galaxies with a polarization fraction of 2 to 3% (details on these tests are given in the supplementary text). This is in line with the expectation that normal galaxies are not strongly polarized, with fractions of less than 10%. Looking at the current largest catalog of polarized sources (49), the mean polarization fraction of all detected sources is only 6%, and only 1.7% of the sources have a fraction of >20%, or a sky density of highly polarized point sources of just 0.02 per square degree. Thus, with the total intensity alone (as with the original low-frequency detections), it was difficult to rule out galaxies as a major contributor; however, the high degree of polarization strongly favors a nongalactic origin. No other circumstance produced the specific pattern of a ring of polarized emission around the source or cluster. Other sources (like galaxies) might mimic or contribute to the emission seen in the cluster center. However, it would be expected that only accretion or merger shocks can produce the observed peripheral and ring-like highly polarized emission.

The results seen from stacking single clusters fit well in line with the theory, although this emission has not been observed before at these faint levels, at these distances from the clusters, or in low-mass clusters. From a compilation of known cluster relics (11), relics tend to be found at distances around 0.5 to 2 Mpc from the cluster centers; in double relic systems, the average distance is 1.03 Mpc (12). They also tend to be strongly polarized (20 to 30%) (10–12), while the central part of the clusters may be filled with diffuse emission from radio halos, which, to date, have not been detected in polarized intensity due to the high levels of turbulence in the central regions of the clusters causing depolarization. Individual detections of relics have only been made in approximately 100 clusters, all with higher masses. Thus, the stacking of the single LRGs or clusters and the clear highly polarized ring of emission provide strong evidence that this type of emission persists even in smaller cluster or group systems below previous detection thresholds.

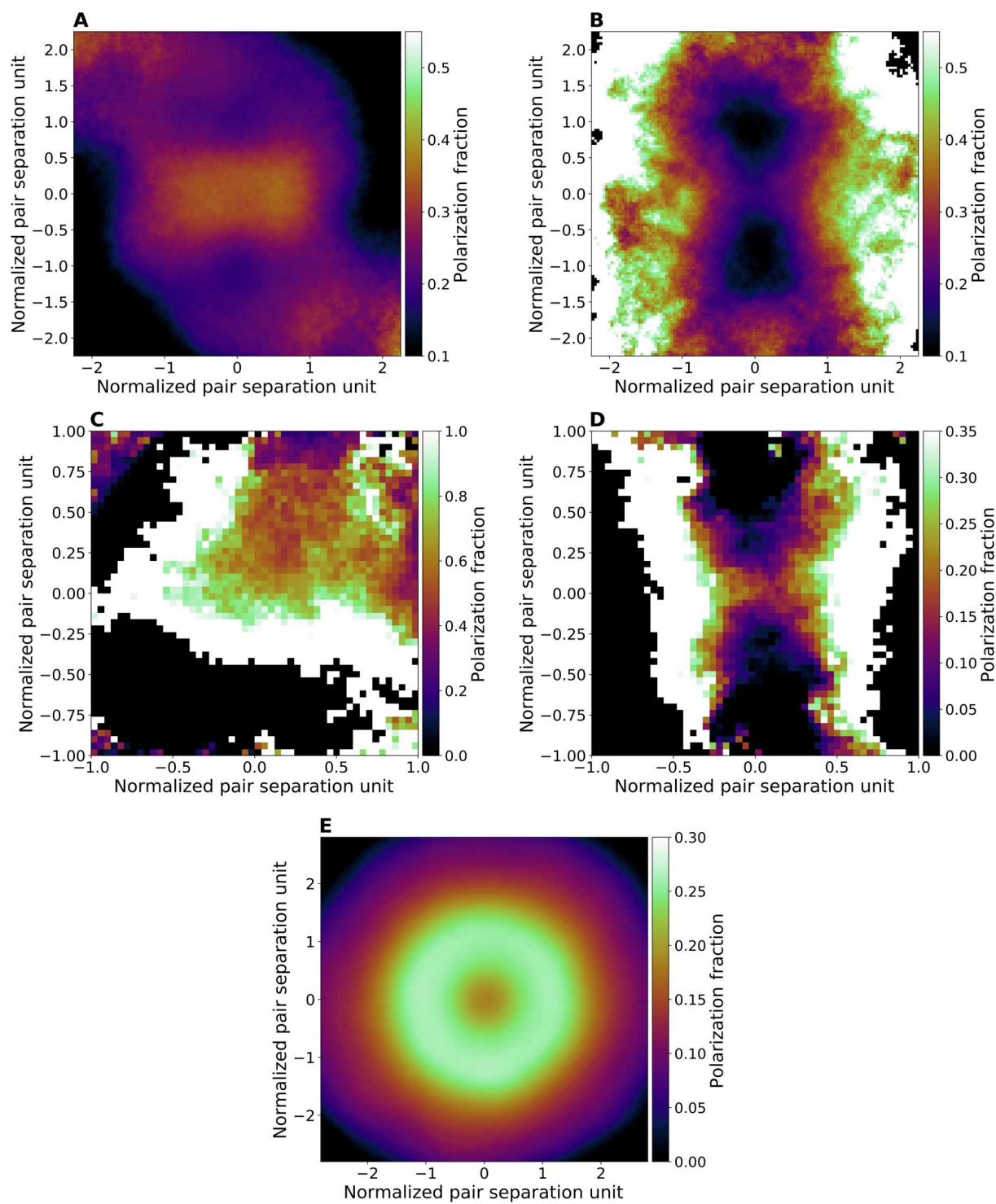

**Fig. 4. Polarization fractions of the 1.4- and 30GHz stacks and residuals.** (**A** and **B**) The 1.4- and 30-GHz stack polarization fractions, respectively. (**C** and **D**) Polarization fractions from the stack residuals. (**E**) The polarization fraction for the single cluster stacking on the 1.4-GHz data. All are zoomed in from the original maps to avoid the noisy regions away from any signal.

Simulations of structure formation also predict a larger population of shocks, with a Mach number overall increasing with the radius, as a result of ubiquitous supersonic accretions at the periphery of clusters (50, 51). These shocks are expected to contribute to some level of synchrotron radio emission, typically below the detection level of known radio relics (5, 52). Until now, radio emission from these shocks has never been observed. At these radii, the predicted amplitude of the radio emission associated with these shocks is quite uncertain, owing to the poorly constrained amplitude of magnetic fields, which may scale with the actual scenario for magnetogenesis. For example, by simulating the cross-correlation signal between the halo distribution and the diffuse synchrotron radio emission from structure formation shocks, recently, it has been shown that the correlated radio signal is quite different, depending on whether magnetic fields have been seeded by primordial or astrophysical mechanisms, out to several megaparsecs from the center of halos (53).

Direct detections of radio emission from filaments remain challenging. An initial detection of a signal between stacked pairs at low radio frequencies has been made in our earlier work (27). There have been detections of intercluster bridges (22, 23, 25), as well as intracluster bridges such as seen in the Coma cluster (24), although none of these included any polarization detection or analysis and are primarily detected at low frequencies.

Recent work (34) used Faraday RMs from the Low-Frequency Array (LOFAR) Two-meter Sky Survey Data Release 2 (LoTSS DR2) (54, 55) to estimate magnetic field values in filaments of 32 ± 3 nG, consistent with the findings from the low-frequency stacking of our group (27), which estimated magnetic field values from the low-frequency stacking in the 40- to 60-nG range (which remain consistent with this work with the α ≃ −1 spectral index, and the equipartition equation breaks down for α > −1.). If the emission detected here is from filaments, as assumed from the LOFAR RM work, then the fact that the line-of-sight magnetic field strength (from RMs) is roughly equal to the plane-of-the-sky magnetic field strength (from synchrotron emission) again implies a low level of turbulence in filaments.

The closest comparisons to this work, in terms of frequency range and polarization measurements, come from studies of cluster relics. There have only been a handful of relics detected and studied at high (>3-GHz) frequencies. The two most well-studied relics, with the most high-frequency data (including the only relics studied at 30 GHz), are the "Sausage" and "Toothbrush" relics. These two, with a handful of others at mid to high frequencies, have been found to have power-law spectra with spectral indices ranging from −0.9 to −1.44 and polarization fractions in the range of 13 to 40% (56–59).

While few in number, the findings from studies of cluster relics agree well with the findings in this work. The excess between the stacked connected clusters has a spectral index in the range of −0.85 ± 0.1 to −1.05 ± 0.15, consistent with typical values of ∼ −1 from relic compilations (10, 12) (and references therein). In another compilation of relics (11), the minimum reported integrated flux density of a known relic at 1.4 GHz was 2.4 millijansky (mJy) (with a mean of the reported relics being tens of millijanskys). Here, the intercluster signal at 1.4 GHz, converted to jansky, has an average value of roughly 1 mJy (although as an average over a large angular area). Recent observations are pushing the bounds to discover previously unidentified low–surface brightness relics, such as the LOFAR observation in (29) that found a radio relic with an average surface brightness of 13 mJy arc min$^{-2}$ at 144 MHz (or less than 1 mJy arc min$^{-2}$ when scaled to 1.4 GHz).

The polarization fractions measured here, both for the cluster outskirts of the single cluster stacking and for the intercluster region of pairs, have values of ≥20% (see Fig. 4 and Table 1). Both measurements are in line with the findings from cluster relics. High degrees of polarization imply a highly ordered magnetic field. For relics, this should result from the compression, which aligns unordered magnetic fields with the shock plane. The high degree of polarization that we see here leads us to the conclusion that we detected shocked emission with our stacking analysis. This is the first detection of shocked emission in filaments and low-mass clusters.

To see how these findings compare with theoretical predictions, we used the latest version of Eulerian cosmological simulations with extragalactic magnetic fields. The simulation covers a comoving volume of $100^3 Mpc^3$ with a constant spatial resolution of 41.6 per cell (comoving), and it evolved a simple uniform magnetic field of $B_0 = 10^{-10}$ G comoving, in a ΛCDM cosmology and using ideal magnetohydrodynamics. The total radio intensity of this simulation has already been computed (5). However, we computed the Stokes $Q$ and $U$ parameters of this emission at 1.4 GHz and at the average redshift of the real sample of cluster pairs ($z = 0.08$).

Thus, we estimated the polarized intensity at 1.4 GHz, which is the first prediction of the polarized emission in such large simulated cosmological volumes. We added noise and convolved the simulations by a beam matching the size of the GMIMS 1.4-GHz data. Using a catalog of halo pairs in the simulation, a similar stacking procedure was repeated on the simulated total and polarized intensity images. These can be seen in Fig. 5.

The resulting stacked images are very similar in appearance to that of the real data images (Figs. 1 and 2), as well as there being a residual central excess in both total and polarized intensity representing the average from intercluster or filament emission. Before any model subtraction, the total intensity shows two peaks at the halo locations, as is seen in the real data images. Furthermore, it shows a ring of diffuse emission around the cluster outskirts. This ring is not seen in the real data. However, the simulations do not include any other radio sources (e.g., radio galaxies or radio halos), underestimating the emission in the central cluster regions. Hence, the brightness of the central cluster regions in the real data is likely just brighter in comparison to the outskirts where the shocked ring resides. The polarized intensity from the simulations looks visually similar to the real polarized intensity images. The central halo regions are depolarized, and they are surrounded by rings of emission. Furthermore, the emission peaks at the center where these rings overlap.

The mean of the central filament region of the simulated stacking is 8.2 μK in total intensity and 2.2 μK in polarized intensity. This is a factor of approximately 10 to 30 smaller in total intensity compared to the intercluster signal in the real 1.4-GHz data (depending on which box is used for measurement). In polarized intensity, the mean is a factor of 37 smaller than the 1.4-GHz polarization signal. The mean fractional polarization from the simulated regions after beam convolution is approximately 25%, a decrease of more than half from before the beam convolution (original values of ~70% in emitting regions of the raw simulated images). This value is in line with what is seen in other predictions from simulation in terms of beam depolarization from cluster shocks when the beam size is larger than the physical scale of the shocks (60). This value of 25% is also similar to what is seen in the real data stacking (with values ranging around 60% at 1.4 GHz or closer to 20% at 30 GHz; see Table 1).

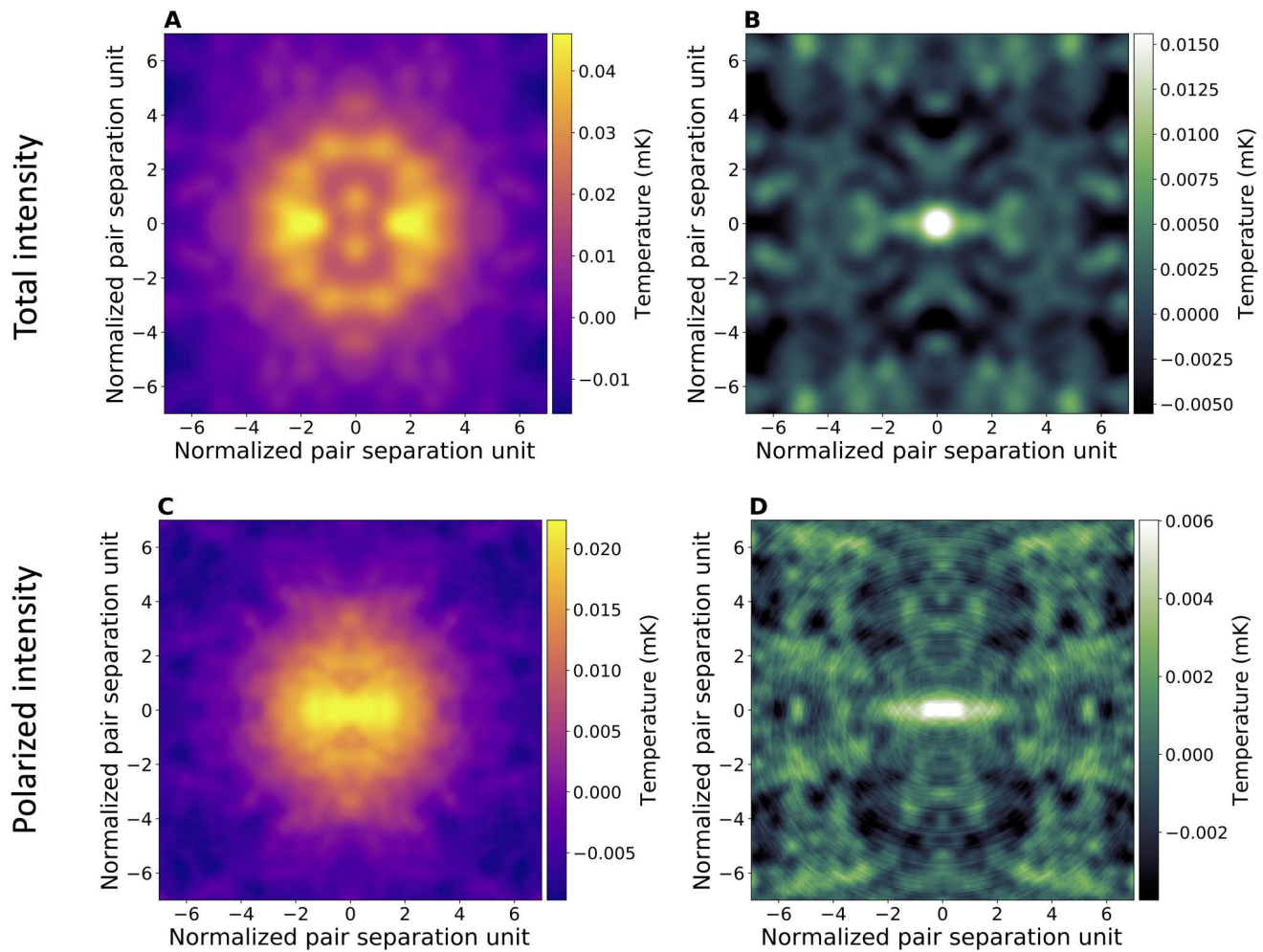

**Fig. 5. Stack images and residual images from stacking on simulations.** (**A**) and (**B**) are the stacked image and residual image from the total intensity, while (**C**) and (**D**) are the stacked image and residual image from the polarized intensity. These are from the simulation with no rescaling and $B_0 = 0.1$ nG and are the averages from 50 resampled iterations with different noise arrays.

Since the simulated values came out below the detected amplitudes, we wanted to see how much stronger the simulated magnetic fields would need to be to match the data. The stacking of the simulated data was repeated assuming larger magnetic field values. Therefore, we recomputed the unpolarized and polarized emission in the simulation assuming magnetic field values that were rescaled by factors of 2, 3, and 4. The scaling of the synchrotron emission after these variations is very well approximated by $\propto B^2$, considering that the magnetic fields in simulated cluster outskirts are always smaller than the microwave-equivalent magnetic field [$B_{CMB} = 3.27$ μG$(1 + z)^2$] and that the emission from structure formation shocks has the flattest possible radio spectrum (with the rare exception of the steeper emission expected from internal merger shocks, possibly related to radio relics). The final stacked values do increase approximately like the scale factor squared. The largest scale factor ends up with the filament region being 1.2 times larger than that measured in the total intensity 1.4-GHz data, with the polarized intensity still a factor of nearly 2.5 smaller than the 1.4-GHz polarization signal. However, taking into account the uncertainties in the data and the possibility of imperfect subtraction of the cluster component in the central regions, it seems that a scale factor of 3 to 4 produces simulated stacks that not only visually resemble the data but also have comparable amplitudes. On the other hand, the shock acceleration efficiency for relativistic electrons used here is rather conservative, as it implies a ~1% overall conversion of shock kinetic power into cosmic ray electron power. Recent works using particle in cell simulations, however, tentatively produced a factor ~10 larger efficiency for the strong shocks responsible for the radio emission from our simulated cosmic web (61). A larger acceleration efficiency would mean that the magnetic field strengths would not need to be increased by as much.

Using the same base simulation with a uniform primordial magnetic field, a recent work using LOFAR RMs (*62*) was able to rule out the simple uniform primordial magnetic field model assumed in the cosmological simulations we analyzed here, based on the evolution of RMs in filaments in favor of a stochastic primordial model. A variety of primordial model, with an initial power-law distribution of tangled magnetic fields, has been shown to be potentially compatible with available limits from the modeling of high-order multipoles in the cosmic microwave background (*63*). For example, stochastic primordial models with a "blue" spectrum actually can have a larger magnetic field energy than uniform primordial ones, distributed into their fluctuating field, yet they can be equally compatible with low-redshift radio observables (*64*). On the other hand, when comparing the uniform versus stochastic primordial models for long lines of sight probing to high redshift ($z = 3$), the different evolutionary trends of uniform and tangled magnetic fields allow further narrowing down the models for cosmic magnetism allowed by radio data [see fig. 6 in the work of Carretti *et al.* (*62*)]. Thus, the stacking analysis here, which used the $z = 0$ (then scaled to $z = 0.08$), may still be compatible, or perhaps more so, with the stochastic primordial model as well. However, the region from which the stacked signal comes from is across the overdensity regime in which signatures of primordial magnetism may mix with the outcome of compression and magnetic field amplification (at least to some extent) driven by structure formation processes, hence magnetic fields likely amplified by small-scale dynamo in the intracluster/intragroup medium. Therefore, the careful modeling of the stacking of radio emission from cluster pairs in more sophisticated models of cosmic magnetism must be deferred to future work. Until then, this work provides the only observational evidence for polarized accretion shocks in filaments from a primordial magnetogenesis model.

While the previous total intensity low-frequency detection was an important first step and is strengthened by the total intensity detections in this work, it did leave room to question whether the emission was from diffuse emission in filaments (versus just galaxies or other sources) and did not answer the question of the type of particle acceleration mechanism at play. Here, the polarization signal, both in its strength and spatial shape, in combination with the total intensity detections, really confirms that what we are seeing is the predicted emission from the formation and growth of the large-scale structure of the Universe.

## MATERIALS AND METHODS

Throughout this work, we adopt a ΛCDM cosmology from (*65*) with $\Omega_m = 0.3075$, $\Omega_\Lambda = 0.6910$, and $H_0 = 67.74$ km s$^{-1}$ Mpc$^{-1}$, and we define the spectral index, $\alpha$, such that the observed flux density $I$ at frequency $\nu$ follows the relation $I_\nu \propto \nu^{+\alpha}$, while the spectral index in units of brightness temperature is defined as $\alpha_T = \alpha - 2$.

### Data

We make use of data from the GMIMS (*47*) HBN. The survey covered 72% of the sky from declination of −30° to +87° at all right ascension. The survey was carried out in both total intensity and linear polarization ($I, Q, U$) over the range of 1280 to 1750 MHz (with a channel width of 236.8 kHz), using the John A. Galt Telescope (diameter, 25.6 m) at the Dominion Radio Astrophysical Observatory. The spectral and Faraday depth cubes are publicly available. The Faraday depth, ϕ, is defined as

$$\phi = 0.81 \int_{l=0}^{l=L} n_e \, B_\| \, dl \text{ rad m}^{-2} \quad (2)$$

where the integral is along the line of sight from the source at $l = 0$ to the observer at distance $L$, $n_e$ is the electron density, and $B_\|$ is the magnetic field strength along the line of sight (or the parallel component). If the polarization is expressed as an exponential vector($P = p \, e^{2i\chi}$), where χ is the polarization angle, then when integrating over all possible Faraday depths

$$P(\lambda^2) = \int_{-\text{inf}}^{+\text{inf}} F(\phi) \, e^{2i\phi\lambda^2} d\phi \quad (3)$$

where $P(\lambda^2)$ is the complex observed polarization vector. By taking the Fourier transform, one can obtain a complex Faraday depth cube, $F(\phi)$. Every spatial pixel in the cube gives a spectrum that describes how the amplitude of the polarization vector changes as a function of Faraday depth. For this work, we find the peak polarized intensity image. The value at each pixel in this image is given by the peak value of $F(\phi)$. We also average the Stokes $I$ over the spectral cube to make a central frequency Stokes $I$ image.

In addition to the GMIMS data, we also make use of the Planck 30-GHz maps (*48*). At the high frequency of the Planck data, the polarized emission vector undergoes negligible Faraday rotation in contrast to the lower-frequency GMIMS data, where the Faraday rotation is more than 400 times greater. For example, at typical Galactic Faraday depths (∼ |150| rad m$^{-2}$) and over a path length of a few tens of kiloparsecs (the path through the Galaxy), the rotation at 30 GHz is expected to be less than 1°, whereas at 1.4 GHz, the polarization vector would undergo nearly a full 360° rotation. This Faraday rotation may lead to depolarization particularly for lines of sight where rotation and emission are occurring together (such as diffuse emission from different distances within the Galaxy). This depolarization effect should be less of an issue for emission from a distant coherent polarized source; however, it may result in a decrement in the observed signal in the lower-frequency observation. However, given the power-law spectrum, the polarized emission should be substantially brighter and may be easier to detect. Thus, we perform our analysis at both frequencies.

The Planck data are all sky maps in Stokes $I$, $Q$, and $U$. We generated the polarized intensity map, $P$, as

$$P = \sqrt{Q^2 + U^2} \quad (4)$$

The effective beam width at 30 GHz is approximately 32 arc min. Although in this work, the Planck maps were converted to a lower resolution of approximately 1° to reduce noise.

All images were in the Healpix format (*66*), with the GMIMS data having $N_{side} = 256$ and the Planck maps having $N_{side} = 128$. The Stokes $I$ and polarized intensity images are shown in fig. S1, with the image properties given in table S1.

As with (*27*), we use LRGs as a tracer of large-scale structure. We use the SDSS Data Release 7 LRG catalog (*46*) (with a total number of LRGs in the catalog of ∼1,400,000, all with photometric redshifts). The presence of an LRG does not guarantee a group or cluster, and thus, there will be a portion of the pairs used that do not contribute to any detected signal or may dilute a true astrophysical signal. However, using LRGs as tracers provides a sufficiently large sample that the majority should still be representative. Cross-matching the SDSS groups and cluster catalog (67) with the LRG catalog found that 86% of the groups and cluster had an LRG match within 1 Mpc.

From all of the computed pairs with a comoving separation of $1 \leq \Delta R \leq 15$ Mpc, we select only those with angular separations of $50 \leq \Delta \theta$ [arc min] $\leq 180$. The lower limit is set so there is at least one beam width between the pairs. The upper limit is set to keep from having too large of a range of $\Delta \theta$s interpolated onto the same grid. From this, we have a total of $N = 612,025$ LRG pairs. We note that the original work in (27) used 390,808 pairs, whereas the total number that met the original criteria was 1,198,412. Some pairs with larger physical separations (but still within the criteria) were missed in the original sample selection; this is discussed further in the Supplementary Materials.

The separation distributions and the redshift distribution are shown in fig. S2. The average angular separation of the pairs is $\langle \Delta \theta \rangle = 101$ arc min, with the average comoving physical separation $\langle \Delta R \rangle \simeq 11.4$ Mpc. The average redshift of the pairs is $\langle z \rangle = 0.08 \pm 0.01$, with a maximum redshift of $z_{max} = 0.255$. The lower limit on the angular separation filters out the higher-redshift pairs that were included in (27).

### Stacking method

We follow the method for stacking laid out in similar filament stacking works (36–38), with details also in (27). For each pair, we follow the method laid out in these previous works and make a 2D cutout or stamp around the two LRGs. This cutout is transformed onto a normalized 2D image coordinate system ($X$, $Y$), with one LRG placed at (0, −1) and the other placed at (0, +1). This requires both a scaling of the pixel sizes and a rotation to align the two LRGs along the vertical axis. This scale factor and rotation angle will be different for each individual cutout. The transformation from sky coordinates to a normalized grid is applied to each cutout map. Then, the average background signal in the cutout is estimated. The mean signal in the annular region $9 < r < 10$, where ($r^2 = X^2 + Y^2$), is subtracted as an estimate of the local background. This is done to ensure that the mean value of each cutout going into the stack is approximately zero. After the transformation and background subtraction, each cutout is added to the previous cutouts. A weight image is also made for each cutout. The weight consists of ones and zeros: one where a pixel is valid (i.e., not NaN or inf values) and zero if the pixel is invalid. The weight maps are also summed. The final stacked images are then made by dividing the image sums by the weight sums. Through the regridding and rescaling process, the surface brightness of the emission is conserved.

A model for the emission outside the region between the LRGs is derived by assuming radial symmetry around each LRG excluding the region between the pair, and then the model for each individual LRG is added together. The model for the two LRGs is then subtracted from the stacked image to get a residual image. Any excess emission between the two LRGs would be apparent in this residual. Further details on this procedure are given in (27) or (37).

Concerns about the assumption of symmetry for this method of modeling the cluster contribution were addressed in the low-frequency radio stacking paper (27), as well as here by stacking on single clusters. It is shown that stacking on single LRGs or clusters does produce a radially symmetric signal if enough samples are included. We note that for truly connected cluster pairs, it is likely that there would be at least slightly higher signal in the direction of the pair or connection. This may thus contribute to any signal found in this intercluster region by using the above method for subtraction. However, as this paper discusses emission from intercluster regions or filaments, as well as cluster outskirts, that emission would still count as detection of emission from large-scale structures.

### Resampling and null tests

To test whether any detected emission between the cluster pairs may be coming from filaments, we performed a null or control test. For this, we selected $N = 612,025$ LRG pairs with the same angular separation distribution as the physical LRG pairs but with physical separations much larger than 15 Mpc. Given that the $\Delta z$ between the two LRGs must be larger for a larger physical separation, these null pairs cannot have the same redshift distribution as the physical pairs. However, at least one LRG in the pair is selected to have a similar redshift as a corresponding physical pair of LRGs, thus giving at least one LRG of the pair approximately the same redshift distribution. We would not expect any strong or filamentary excess between these unconnected pairs, beside the possible residual emission from large-scale structures of the cosmic web, randomly projected along the line of sight. On the basis of cosmological simulations, we can expect this signal to be ≤10 times lower than the signal from filaments physically joining pairs of interacting halos (68). Stacks were made in the same way as the physical pairs, with the same number of pairs. We performed 500 null tests for each radio map, with randomly chosen LRGs for each test.

To test the scatter of the physical LRG pairs, we bootstrapped the sample. We selected 75% of the total number of pairs, randomly chosen with replacement, and repeated the stack for each map. We performed this bootstrapping 500 times for each map. From this, we can look at the scatter of any detected signal. This also helps to alleviate any effect of bright outliers in the stack.

To check the overall chance occurrence or variance of the stacks, we chose $N_{LRG}$ pairs of random sky positions with the same angular separation distribution of the physical pairs and stacked on these random sky positions. This was repeated 100 times. All of the images shown are the averages taken across the different iterations of the resampling.

### Uncertainty

To obtain the uncertainties listed in Table 1, we measured the mean in the different box sizes at all positions in the residual maps. We did this by convolving the residual maps with a mean filter having the shape of the bigger boxes, as well as the optimized total intensity boxes. We then looked at the distributions of pixel values from these convolved maps everywhere outside the central box regions and measured the SDs. This tells us, given the noise per pixel and the correlation scale of the noise, how common it is to find another region in the same map that would yield the same results or higher than is reported in the central region. The mean filter–convolved maps for all the connected pair residuals are shown in fig. S3 for the 1.4-GHz data and in fig. S4 for the 30-GHz data, along with the pixel histograms compared with the measured central values. Using the total intensity boxes, the central excesses in total and polarized intensities at both frequencies have signal-to-noise ratios of 2.8 to 4. In all cases, the values of the central excesses for connected pair maps are always higher than when compared to the unconnected test cases (comparing Table 1 to table S2).

When computing the polarization fractions, the uncertainty from the total intensity and polarized intensity is propagated through, but the polarization fraction uncertainty also takes into account any covariance. To estimate the covariance between the total and polarized intensities, we choose $N_{pairs}$ random positions on the sky and stacked at the same random positions for both maps and repeated this 50 times. For each random stack, we computed the variance for each map and the covariance between the total and polarized intensities and used the average of the 50 covariances. The average correlation coefficient between the 1.4-GHz polarizations is 0.35, and the average for the 30-GHz maps is 0.45. The polarized fraction uncertainties in Table 1 are given with and without taking into account this covariance estimate.

### Single clusters

The appearance of the polarized intensity pair stacks differs markedly from that of the stacks seen in total intensity. In the case of the GMIMS polarization data, a single central bright source is seen in the stacked image rather than the two bright sources at the positions of the clusters as is seen in the total intensity data. In the Planck polarized intensity stacks, there is a centrally located bright emission, and at the positions of the two clusters, there appears to be a deficit of signal.

To investigate the nature of this deficit, we performed tests of stacking on a single cluster. This was performed in the same manner as the pair stacking (i.e., same sample and same grid); however, for each real cluster pair, one of the clusters was replaced by a randomly selected position at the same angular distance away. This was performed using the sample of connected clusters, as well as the more general sample of clusters that the unconnected pairs were chosen from. This test was performed 100 times for each map (e.g., 100 stacks for the GMIMS and Planck connected pairs and 100 for the unconnected clusters). This was done separately for clusters at each of the transformed coordinate positions such that there were now tests where the clusters were at (0, +1) and a random position at (0, −1) and the reverse.

The averages of 100 of these stacks for the polarized intensity maps are shown in fig. S5. From these, we can see that the individual clusters are not just compressed circular sources at the positions of the LRGs. In both cases, they are much more diffuse or spread out. In addition, in both cases, the exact position corresponding to the LRG is not the peak. This is quite noticeable in the case of Planck, which shows a decrement, or hole, around the position of the LRG, forming a ring of signal around the LRG position. This is also the case in the GMIMS data, although to a much lesser extent, but the signal does decrease slightly at the LRG position.

When the top and bottom single cluster stacks are averaged together, the resulting signal patterns closely resemble that which is seen in the pair stacking (Fig. 2, A to D). Thus, although none of the polarization pair stacks look like two distinct clusters, that is the case. We can also subtract off these coadded single stacks from the original pair stacks; this is shown in fig. S5. In this case, the connected pair stacking still shows a central excess in the regions between the two clusters; however, the unconnected tests no longer show any excess signal in the central region. This implies that the connected pairs are simply the addition of stacking two independent clusters, whereas in the physically related case, the two independent cluster stacks added together cannot account for all the emission seen in the region between them.

In this test, the original stacking setup was maintained such that the "pair" was made up of a single LRG and a randomly chosen position at the same distance away from it as the "real" physical pairing. This maintained the same scaling and rotation to compare with the pair stacking. However, seeing that there was detectable emission in the single cluster stacks, we decided to investigate this further. We chose all the LRGs from the original catalog that have $0.025 \leq z \leq 0.3$ ($N = 280,050$) and stacked with those positions as the central positions. This allowed for a scaling based solely on redshift such that the pixel units have a physical scale in megaparsecs. We performed 100 iterations of this test for each map using a randomly selected subset of the single LRGs available in each iteration. The results from the averages of these 100 stacks are shown and discussed in the Results section.

### Simulations

Similar to our previous paper (27), we have used the cosmological simulation of extragalactic magnetic fields presented in (5), which was produced using the magnetohydrodynamic code ENZO (www.enzo-project.org). The simulation domain covers a comoving volume of $100^3$ Mpc$^3$ with a uniform grid of $2400^3$ cells and $2400^3$ dark matter particles, giving a constant spatial resolution of 41.6 per cell (comoving). The simulation was initialized with a simple uniform magnetic field of $B_0 = 10^{-10}$ G comoving at $z = 45$, and the adopted cosmological parameters for the $\Lambda$CDM model are $\Omega_{BM} = 0.0478$, $\Omega_{DM} = 0.2602$, $\Omega_\Lambda = 0.692$, and $H_0 = 67.8$ km/s per megaparsec.

This simulation does not include radiative gas cooling, star formation, or feedback processes on the premise that such processes are not crucial to model the production of large-scale radio emission in the shocked cosmic web. The latter is computed, for this updated modeling, with a more advanced approach than in our paper I or in previous works. For recent applications of this large simulation to the modeling of low-frequency radio surveys, see also (69) and (70).

To compute the radio emission from the shocked cosmic web, we followed the approach of (71). Here, we summarize the main steps of the computation, and we point to (71) for details.

Using a velocity jump–based shock finder, we searched for shock waves in the simulation. Following equations 1 and 2 of (71), we computed the radio emission at a distance $r$ from the shock front as the convolution of the electron energy spectrum at distance $r$ and the modified Bessel function: $I(\nu, r) \propto \int n_e(E, r) F(E) dE$. Similarly, we computed the perpendicular and parallel part of the emission as $I_\perp(\nu, r) \propto \int n_e(E, r)[F(E) - G(E)]dE$ and $I_\parallel(\nu, r) \propto \int n_e(E, r)[F(E) + G(E)]dE$, respectively. Using these quantities, we computed the corresponding polarized emission as

$$I_{pol}(\nu) = \frac{\sum_{los} I(\nu) \Pi \exp[2i(\epsilon_{int} + RM\lambda^2)]}{\sum_{los} I(\nu)} \quad (5)$$

following (72). In the equation above, $\Pi = (I_\parallel - I_\perp)/(I_\parallel + I_\perp)$ is the intrinsic degree of polarization, $\epsilon_{int}$ is the intrinsic angle of polarization, $\lambda$ is the wavelength corresponding to the observing frequency ν, and RM is the rotation measure. The latter is computed as

$$\mathrm{RM}(x) = 812 \int_0^x \frac{n_e}{10^{-3}\mathrm{cm}^{-3}} \frac{B_\parallel}{\mu\mathrm{G}} \frac{\mathrm{d}l}{\mathrm{kpc}} \qquad (6)$$

The RM at a distance $x$ from the observer depends on the thermal electron number density, $n_e$, and the magnetic field parallel to the line of sight, $B_\parallel$.

Equation 5 is directly connected to the Stokes parameters. Here, the denominator corresponds to Stokes $I$, while the real part and the imaginary part of the numerator correspond to Stokes $Q$ and $U$, respectively.

All physical quantities that are needed to compute the polarized emission are directly taken from the simulation. This leaves the acceleration efficiency ξ [see equation A10 in (71)] as the only free parameter in our model. ξ directly enters the normalization of the electron spectrum, and here, we use ξ = 0.02, which has proven to be a reliable choice in the past (71, 73). A zoom-in of the simulated Stokes images ($I$, $Q$, and $U$) for a rectangular ≲70 Mpc × 30 Mpc × 100 Mpc region of the ENZO cosmological simulation used to produce the theoretical expectations from the shocked cosmic web is shown in fig. S6. The last panel gives the polarization fraction of the emission, also including the effect of Faraday rotation along the line of sight, using the 3D magnetic field information simulated at run time. Figure S7 shows the combined total intensity radio emission along with x-ray emission and the magnetic field vectors. The volume is placed at $z = 0.08$, and the resolution of cells is 41.66 kpc (comoving). The image shows the formation of sharp emission features surrounding the periphery of cosmic structures (filaments and outskirts of halos), where strong shocks continuously form. Such features are often associated with the strongest degree of polarization in the model, while the emission from the multitude of weaker (merger) shocks in the innermost regions of halos overall shows a smaller degree of polarization, owing to the depolarization by intracluster magnetic fields.

Taking the output of the simulations, we scaled the emission to a set redshift equal to the mean of the LRG pairs sample, $z = 0.08$. We converted the units to jansky and then added random Gaussian noise to the images. We then convolved the Stokes $I$, $Q$, and $U$ images by a 40–arc min beam to match that of the GMIMS data. After convolution, we generated a polarized intensity $P$ image from the beam-convolved Stokes $Q$ and $U$ maps. We used the same catalog of mass halos, described in (27), as clusters that are within 15 Mpc of each other ($N_{\mathrm{halos}} = 2036$). With this sample of physically nearby halos, we repeated the same type of stacking procedure (e.g., same pixel grid and scaling as the real data) and stacked the pairs of halos in both the total intensity and polarized intensity. We created a final stacked image from the simulated pairs in both total and polarized intensity.

Given that the number of simulated pairs is much less than the real number of LRG pairs, we created 50 stack realizations from the simulations by repeating the stacks with randomly generated resampled (with replacement) simulated halo pairs, as well as flipping the orientations of the pairs, and generating different noise realizations with each iteration. The noise added was approximately a factor of 10 to 20 lower than the noise in the real GMIMS maps, as $\sqrt{N_{\mathrm{halos}}}$ is 17 times smaller than $\sqrt{N_{\mathrm{LRG}}}$, meaning the noise would be reduced by a smaller amount from stacking of the halo pairs. We then took the average of the 50 simulation stack realizations and performed a similar model of cluster emission and subtraction to obtain a residual filamentary excess; examples of these can be seen in Fig. 5. This was repeated using the scaling factors of 2, 3, and 4 for the magnetic field strengths in the simulations to produce a scaling relation of $B^2$ on the synchrotron emission. The measured values for the filamentary excess emission in the total and polarized intensity simulation stacked images are presented in table S3.

One thing to note here is that the simulations only include radio emission from shocks. They do not include emission from galaxies or other sources of diffuse emission (e.g., radio halo–like emission or Fermi-II processes), so it is likely that this is something of a lower limit, or less than the data stacking, inherently because of this reason. Therefore, while any filamentary emission, and the outskirts of clusters, is thought to be dominated by shocks to produce diffuse emission, there is likely more that contributes to the data stacking.

## Supplementary Materials

**This PDF file includes:**
Supplementary Text
Figs. S1 to S10
Tables S1 to S3
References

**Acknowledgments:** We would like to acknowledge the help of H. Tanimura, J. Stil, and A. Thomson. **Funding:** J.W. acknowledges the support of the Natural Sciences and Engineering Research Council of Canada (NSERC) through grant RGPIN-2015-05948 and of the Canada Research Chairs program. The Dunlap Institute is funded through an endowment established by the David Dunlap family and the University of Toronto. F.V. acknowledges financial support from the Horizon 2020 program under the ERC Starting Grant MAGCOW (no. 714196). D.W. is funded by the Deutsche Forschungsgemeinschaft (DFG; German Research Foundation) (441694982). C.J.R. acknowledges financial support from the ERC Starting Grant "DRANOEL" (no. 714245). The cosmological simulation was run on the Piz Daint supercomputer at CSCS-ETH (Lugano), under project s1096. We acknowledge the Gauss Centre for Supercomputing e.V. (www.gauss-centre.eu) for supporting this project by providing computing time through the John von Neumann Institute for Computing (NIC) on the GCS Supercomputer JUWELS at Jülich Supercomputing Centre (JSC), under project no. hhh44 and TuMiB (PI: D.W.). **Author contributions:** T.V. was the lead on this project carrying out the methods and analysis and writing of the majority of the text. J.W. was involved in the inception of the idea, providing valuable feedback about the analysis and techniques and interpretation as well as edits to the text. F.V. and D.W. supplied the cosmological simulations, as well as the text describing them. F.V. and D.W. also provided text and feedback regarding the interpretation. G.H. and C.J.R. provided feedback on the text and on the interpretation. **Competing interests:** The authors declare that they have no competing interests. **Data and materials availability:** All data needed to evaluate the conclusions in the paper are present in the paper and/or the Supplementary Materials. All data used are publicly accessible. The GMIMS maps can be obtained from www.canfar.net/storage/vault/list/AstroDataCitationDOI/CISTI.CANFAR/21.0003/data/release. The Planck data can be obtained from https://irsa.ipac.caltech.edu/data/Planck/release_3/all-sky-maps/. The Owens Valley Long Wavelength Array map was obtained from https://lambda.gsfc.nasa.gov/product/foreground/fg_ovrolwa_radio_maps_get.cfm. Information on obtaining GLEAM images can be found at www.mwatelescope.org/gleam. The LRG catalog can be downloaded from https://cdsarc.cds.unistra.fr/viz-bin/cat/J/MNRAS/380/1608. The full list of pairs of LRGs used in this work can be accessed at https://zenodo.org/record/7395299. In this work, we used the ENZO code (http://enzo-project.org), the product of a collaborative effort of scientists at many universities and national laboratories. Samples of our simulations are available at https://cosmosimfrazza.myfreesites.net/the_magnetic_cosmic_web. Other python packages used are accessible at https://healpy.readthedocs.io/en/latest/ for Healpy and www.astropy.org for Astropy.

Submitted 2 September 2022
Accepted 13 January 2023
Published 15 February 2023
10.1126/sciadv.ade7233

# Science Advances

**Polarized accretion shocks from the cosmic web**

Tessa Vernstrom, Jennifer West, Franco Vazza, Denis Wittor, Christopher John Riseley, and George Heald

**View the article online**
https://www.science.org/doi/10.1126/sciadv.ade7233
**Permissions**
https://www.science.org/help/reprints-and-permissions